\documentclass[a4paper,fleqn]{cas-sc}

\usepackage[numbers, sort&compress]{natbib}
\usepackage{amssymb}
\usepackage{lipsum}
\usepackage{amsmath}
\usepackage{soul}  
\usepackage{xcolor}

\bibpunct{[}{]}{,}{n}{,}{,}

\def\tsc#1{\csdef{#1}{\textsc{\lowercase{#1}}\xspace}}
\tsc{WGM}
\tsc{QE}
\begin{document}
\let\WriteBookmarks\relax
\def\floatpagepagefraction{1}
\def\textpagefraction{.001}

\shorttitle{}    

\shortauthors{}  

\title [mode = title]{Performance appraisal promotes cooperation in spatial public goods games}  


% \tnotetext[1]


\author[1]{Tianjiao Li}[
 orcid=0009-0002-2081-1065
]

\author[2]{Qin Li}[
    orcid=0000-0001-5480-0992
]
\cormark[1]

\author[1]{Kangxi Zhu}
\cormark[1]

\author[1]{Minyu Feng}[
    orcid=0000-0001-6772-3017
]

\author[3,4]{Manuel Chica}[
    orcid=0000-0002-4717-1056
]

\cortext[1]{Corresponding authors. 
    E-mail addresses: \href{mailto:qinli1022@swu.edu.cn}{qinli1022@swu.edu.cn} (Q. Li), 
    \href{mailto:kangxi2019@swu.edu.cn}{kangxi2019@swu.edu.cn} (K. Zhu).}

\affiliation[1]{organization={College of Artificial Intelligence, Southwest University},
            city={Chongqing},
            postcode={400715},
            country={PR China}}

\affiliation[2]{organization={Business College, Southwest University},
            city={Chongqing},
            postcode={402460},
            country={PR China}}

\affiliation[3]{organization={Andalusian Research Institute DaSCI and Dept. of Computer Science and AI},
            addressline={University of Granada},
            city={Granada},
            postcode={18071},
            country={Spain}}

\affiliation[4]{organization={School of Information and Physical Sciences},
            addressline={The University of Newcastle},
            city={Callaghan},
            postcode={NSW 2308},
            country={Australia}}

\begin{abstract}
In real organizations, performance appraisal serves as an important means of evaluating employees' work outcomes and behaviors, while performance pay directly links compensation to the evaluation results. However, in traditional spatial public goods games, the total payoff of a group is equally distributed among all participants without considering individual differences in contributions, an approach that ignores the heterogeneity of individual efforts. To overcome this limitation, we propose a spatial public goods game model based on performance appraisal, in which individual payoffs are divided into two parts: an equally distributed component and a performance-weighted component. Specifically, each individual receives scores from its neighboring groups, and the individual's reputation is dynamically updated by accumulating these scores, which further modulates the fitness function during strategy imitation. Extensive numerical simulation results demonstrate that the performance appraisal mechanism significantly promotes the emergence of cooperative behavior. The reputation reinforcement mechanism amplifies this positive effect by creating fitness advantages for high-reputation individuals. These findings suggest that incorporating performance appraisal into payoff allocation helps mitigate social dilemmas and promote collective cooperation.

\end{abstract}



% Keywords

\maketitle

% Main text
\section{Introduction}
\label{introduction}

Cooperative behavior requires individuals to voluntarily pay the costs for the sake of shared benefits, a phenomenon widely observed in both natural and human societies~\cite{axelrod1981evolution}. However, Darwin's theory of natural selection, which centers on individuals competing for survival and reproduction, seems to conflict with cooperation for collective interests. Nowak summarized five mechanisms that can promote the emergence of cooperation~\cite{nowak2006five}: kin selection~\cite{eberhard1975evolution}, direct reciprocity~\cite{nowak1993strategy,pacheco2008repeated}, indirect reciprocity~\cite{righi2018social}, network reciprocity~\cite{ohtsuki2006simple}, and group selection~\cite{wilson1975theory}. Under certain conditions, cooperation is an effective strategy for individuals to maximize their own interests. Consequently, understanding the spontaneous emergence and maintenance of cooperation between selfish individuals has become a key topic of interdisciplinary research~\cite{perc2017statistical,li2023open,archetti2011economic,capraro2024language}. To address this issue, evolutionary game theory provides the most effective analytical framework. Researchers have explored various game models, among which the most representative include the prisoner's dilemma~\cite{rapoport1965prisoner,yao2023inhibition,shen2026evolutionary}, the snowdrift game~\cite{doebeli2005models,xiong2024adaptive,wang2024utility}, and the public goods game~\cite{perc2013evolutionary,zhang2025spatial}. Among them, the public goods game is a typical example that describes cooperative behavior in group interactions.

On this basis, the impact of spatial structure on the evolution of cooperation has received extensive attention. In a pioneering study conducted in 1992, Nowak and May introduced a regular lattice into the game framework and explored the role of local interactions in cooperation for the first time~\cite{nowak1992evolutionary}, laying the theoretical foundation for the study of the evolution of cooperation in spatial structures. Since then, researchers have continuously extended this direction, successively investigating the evolutionary mechanisms of cooperation under various topological structures, such as small-world networks~\cite{masuda2003spatial} and scale-free networks~\cite{santos2005scale,yao2026fixed}. Several studies have shown that spatial structure significantly promotes the emergence and maintenance of cooperation through the amplifying effects of local aggregation and reciprocal interactions. Based on spatially structured populations, researchers have attempted to constrain defection through various mechanisms, mainly including punishment~\cite{chen2015competition,gao2020evolution,szolnoki2017second,zhang2025evolution}, reward~\cite{szolnoki2010reward,wu2017impact,zhang2024impact,wang2017role}, exclusion~\cite{szolnoki2017alliance,liu2017competitions} and memory effects~\cite{wang2006memory}. 

Based on the social attributes of humans, the reputation mechanism has been extensively validated by research to effectively promote cooperation~\cite{he2026reputation,zhang2025evolutionary,feng2023evolutionary,pal2022reputation}. It indirectly influences cooperation by recording individuals' historical behaviors and constructing social credits. Building on this insight, many studies have designed various reputation mechanisms to guide cooperative behavior. However, in reality, the reputation evaluation process is much more complex than the simplified models suggest. Different individuals often have heterogeneous evaluation criteria, which can lead to completely different judgments of the same behavior. To this end, relevant research has gradually expanded from first-order norms~\cite{nowak1998evolution,berger2016stability}, which focus solely on behavior itself, to second-order~\cite{dong2019second,dong2019cooperation} and third-order norms~\cite{yang2019evolution}, aiming to capture the various evaluation systems in the real world. This indicates that the reputation mechanism has made significant progress in the study of cooperation evolution and also implies that there remains considerable design space for its role in promoting cooperation.

In response to the free-rider problem in traditional public goods games, existing studies have made improvements from multiple perspectives~\cite{wang2026dynamics,ma2021effect,yang2019reputation,pei2017effects,li2026supervised,wei2026cooperation}. In the classical framework, individual payoffs are distributed completely equally, and rewards or punishments are typically implemented through additional gains or losses. However, in recent years, researchers have begun to focus on the heterogeneity of payoff distribution, attempting to explore new pathways to promote cooperation from the perspective of distribution rules. Peng \textit{et al.} found that in networked populations, heterogeneous distribution based on individual connectivity can promote cooperation~\cite{peng2010promotion}. Huang \textit{et al.}, from the perspective of fairness preference, proposed a dynamic heterogeneous distribution mechanism based on willingness to invest and found that a moderately differentiated distribution can effectively promote the evolution of cooperation in spatial public goods games (SPGGs)~\cite{huang2023evolution}. Gao \textit{et al.} found that distribution policies favoring low-income individuals have positive effects~\cite{gao2019resolving}. Collectively, these studies reveal the promoting effect of heterogeneous distribution on the evolution of cooperation, indicating that structural adjustments to the payoff distribution rules themselves represent a feasible and effective pathway to promote cooperation.

In reality, salaries often consist of guaranteed base income and performance pay. From a management perspective, performance management is defined as a continuous process whose core purpose is not only to evaluate employees' past work behaviors and outcomes but also to guide future behavioral improvements through feedback and development plans~\cite{gruman2011performance}. Performance appraisal, as the core tool to achieve this purpose, is essentially a systematic process that periodically evaluates individuals' work behaviors and achievement results based on specific criteria and methods, and assigns a score to the evaluation result~\cite{denisi2017performance}. Performance appraisal scores not only serve as a record of past performance, but also constitute an important basis for determining salary distribution~\cite{cappelli2018performance}, providing practical support for this paper's introduction of the performance appraisal mechanism into the evolutionary game model.

Based on this observation, our work proposes an evolutionary game model based on performance appraisal. In this model, an individual's payoff is divided into two parts: the first part is equally distributed to guarantee a basic income; the second part is differentially distributed according to the individual's performance appraisal score, where the performance score is dynamically generated through a neighbor mutual evaluation mechanism. The core of this design lies in the fact that the performance score conveys peers' evaluative signals of an individual's cooperative behavior. By influencing both the current payoff and the accumulation of reputation, it incentivizes individuals to adjust their strategies in subsequent rounds, thus forming sustained dynamic incentives. Specifically, in each round of the game, each individual receives scores from its neighbors, which simultaneously affect the update of the individual's reputation and the distribution of payoffs. Furthermore, an individual's strategy update is linked to its current reputation, thus achieving a dynamic coupling between strategy evolution and the individual's reputational status.

The remainder of this paper is structured as follows. Section~\ref{sec:2} provides a detailed description of the proposed model, including the payoff calculation formulas, the reputation update mechanism, and the strategy learning rules. Section~\ref{sec:3} presents and analyzes the simulation results under different parameters and network types. Section~ \ref{sec:4} concludes the research work and discusses future research directions.

\section{Model} 
\label{sec:2}
In this section, we will elaborate on the specific design of the proposed model based on the performance appraisal mechanism, including the payoff calculation for different types of participants, the strategy update rule, and the reputation update rule.

\subsection{Spatial public goods game with performance appraisal}

In the spatial public goods game (SPGG) model, the nodes of the network represent players, and the edges denote the interactive relationships between them, with a total of $N$ individuals. Each player can adopt one of two strategies: cooperation (C) or defection (D). In the initial time step, the two strategies are randomly assigned to each player with equal probability. In each round of the game, all players decide their strategies simultaneously. In this model, each player $i$ participates in public goods games organized by itself and all its direct neighbors. Within each group, cooperators contribute a cost $c$ to the public pool, while defectors contribute nothing. The total contribution of all members of a group is multiplied by a synergistic factor $r$ and then equally distributed among all members of the group, regardless of their strategies. The total payoff of player $i$ in a single round is the sum of the payoffs obtained from all the groups in which it participates, which can be expressed as follows.
\begin{equation}
\pi_i = \sum_{g \in G_i} \pi_i^g
\end{equation}
where $G_i$ denotes the set of all public goods game groups in which individual $i$ participates, and $\pi_i^g$ represents the payoff earned by individual $i$ in a specific group $g$. Consequently, the total payoff $\pi_i$ of individual $i$ is the sum of the payoffs acquired from all the groups to which it belongs.

We propose a performance appraisal payoff distribution mechanism that modifies the traditional SPGG to bring the model closer to reality. Under this mechanism, the total payoff in the public pool is divided into two parts according to a certain proportion: one part is equally distributed among all members of the group, while the other part is allocated to each player based on performance appraisal. Specifically, the payoff $\pi_i^g$ of the player $i$ in group $g$ can be expressed as:
\begin{equation}
\pi_i^g = \alpha \frac{P_g}{ | G_g | } + (1 - \alpha) P_g w_i^g - c S_i
\end{equation}
where $P_g$ denotes the total payoff of the public pool in group $g$ after multiplying the total contributions by the synergistic factor $r$, $ | G_g | $ is the size of group $g$, and $\alpha$ is the proportion of the payoff of the public pool that is equally distributed among all members. The remaining proportion $(1-\alpha)$ is allocated based on performance. $w_i^g$ is the performance weight of player $i$ in group $g$, and the weights within a group satisfy $\sum_{i \in g} w_i^g = 1$. In particular, when all weights are equal, the distribution reduces to the traditional equal allocation, with each weight being $1/ | G_g | $. The term $c S_i$ represents the cost borne by the cooperators, where $S_i=1$ for cooperation and $S_i=0$ for defection.

The performance weight $w_i^g$ is determined by the performance score that each player receives in the current round. Each player first receives a score from each of its direct neighbors, based on the cooperative or defective atmosphere within the neighbor-centered group. This design is motivated by the fact that in real organizations, performance evaluation is typically a multi-source feedback process in which multiple colleagues or supervisors who are familiar with the evaluatee assess the individual from different perspectives. The score given by each evaluator reflects the relative performance of the evaluatee within the evaluator's own working environment. Specifically, if player $i$ cooperates in the current round, the score given by neighbor $j$ is the proportion of cooperators in $j$'s group, which is a positive score; if player $i$ defects, the score is the negative value of the proportion of defectors in $j$'s group, which is a negative score. The final performance score $\varepsilon_i(t)$ of player $i$ is the arithmetic mean of the scores given by all its neighbors:
\begin{equation}
\varepsilon_i(t) = \frac{1}{ | N_i | } \sum_{j \in N_i} \varepsilon_{j \rightarrow i}(t)
\end{equation}
where $N_i$ denotes the set of neighbors of the individual $i$ in the time step $t$, $ | N_i | $ is the number of neighbors and $\varepsilon_{j \rightarrow i}(t)$ represents the score given by neighbor $j$ to the player $i$ in the time step $t$. The specific calculation of this score is as follows:
\begin{equation}
\varepsilon_{j \rightarrow i}(t) = 
\begin{cases}
\dfrac{1}{ | G_j | } \displaystyle\sum_{{{ 
{ k \in G_j} 
}} }^{{{ 
{ S_k = 1} 
}} } 1, & S_i = 1, \\
-\dfrac{1}{ | G_j | } \displaystyle\sum_{{{ 
{ k \in G_j} 
}} }^{{{ 
{ S_k = 0} 
}} } 1, & S_i = 0
\end{cases}
\end{equation}

The performance score obtained by each player in each round is used to determine the allocation weight in each public goods group in which they participate. Due to differences in the distribution of performance scores among members in different groups, the raw performance score of the same player may carry different meanings in different groups. To ensure comparability of performance scores between groups, a normalization process is required. Specifically, within each group $g$, the raw performance scores of all members of that group are linearly assigned to the interval $[0, 1]$. The performance weight $w_i^g$ of the player $i$ in the group $g$ is then calculated according to the following set of equations:
\begin{equation}
w_i^g = \dfrac{\varepsilon_i^{g\ast}(t)}{\sum_{j \in g} \varepsilon_j^{g\ast}(t)}
\end{equation}
where $\varepsilon_i^{g\ast}(t)$ represents the normalized performance score of player $i$ in group $g$ at time step $t$, i.e.,~$\varepsilon_i^{g\ast}(t) = \dfrac{\varepsilon_i(t) - \varepsilon_{\min}^g(t)}{\varepsilon_{\max}^g(t) - \varepsilon_{\min}^g(t)}$, where $\varepsilon_{\max}^g(t)$ and $\varepsilon_{\min}^g(t)$ denote the maximum and minimum raw performance scores within group $g$ at time step $t$, respectively. In the extreme case where all members within a group have identical raw performance scores, the denominator in this expression becomes zero. In this case, we stipulate that the normalized performance scores of all members are equal. The resulting weights satisfy $\sum_{i \in g} w_i^g = 1$, which means that the sum of the performance allocation weights of all members of a group equals unity.

%%%%%%%%%%%%%%%%%%%%
\subsection{Reputation update and strategy evolution}

Within this scoring system, a player's reputation is closely tied to the evaluations of its neighbors. Accordingly, we define the amount of update of a player's reputation in each round as the arithmetic mean of all scores received from its neighbors. The specific reputation update rule is as follows:
\begin{equation}
R_i(t+1) = R_i(t) + \varepsilon_i(t)
\end{equation}
where $R_i(t)$ represents the reputation value of the player $i$ in the time step $t$, and $\varepsilon_i(t)$ denotes the performance score of the player $i$ in the time step $t$, which is the arithmetic mean of all scores received from its neighbors. $R_i(t+1)$ represents the updated reputation value of the player $i$ in the next time step $t+1$. To prevent the unbounded growth of reputation values, we constrain them within the interval $[0, 20]$. If the updated reputation value exceeds 20, it is set to 20. If it falls below 0, it is set to 0.

\begin{figure*}
\centering 
\includegraphics[width=0.9\textwidth]{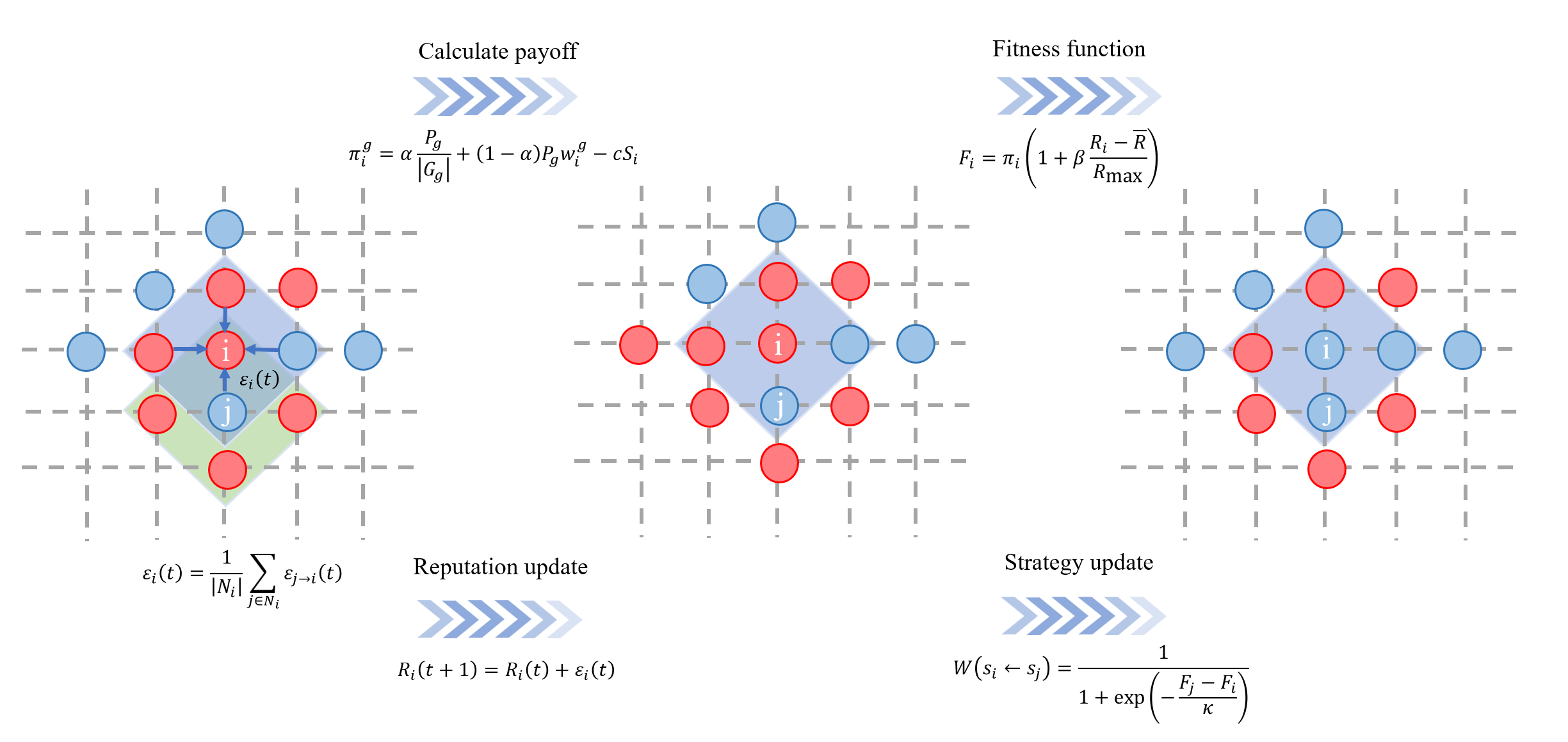}
\caption{\label{fig:1}\textbf{Illustration of the processes of payoff calculation, reputation update, and strategy updating.} Blue and red nodes represent cooperators and defectors, respectively, and edges denote game interactions. The blue shading indicates a game group, while the green shading represents the cooperative or defective atmosphere of a neighbor's group. The model consists of three steps: (i) Scoring: neighbors evaluate the focal node based on the atmosphere of their own groups; (ii) Reputation update and payoff calculation: the focal node updates its reputation based on the average score received from its neighbors and calculates its payoff; (iii) Strategy updating: the focal node imitates the strategy of a neighbor with higher fitness with a certain probability, taking into account both payoff and reputation. }
\end{figure*}

In the real world, an individual's strategy choice is influenced not only by payoffs but also by reputation. To capture this feature, we incorporate the regulatory effect of reputation on payoffs during the strategy update process. Specifically, when a player's reputation is higher than the average reputation of the population, its payoff is amplified. Conversely, when its reputation is lower than the average, its payoff is attenuated. We introduce a parameter $\beta$ $(\beta \geq 0)$ to control the intensity of this reputation regulation. Consequently, the fitness function of the player $i$ is defined as follows, where $\overline{R}$ is the average reputation of the current population, and $R_{\text{max}}$ is the upper bound of the reputation, which is set to 20 in this paper:
\begin{equation}
F_i = \pi_i\left(1+\beta\frac{R_i-\overline{R}}{R_{\text{max}}}\right)
\end{equation}

During the strategy update stage, each player randomly selects a neighbor as an imitation target. The probability that player $i$ adopts the strategy of neighbor $j$ is given by the following equation:
\begin{equation}
W\left(s_i\leftarrow s_j\right)=\frac{1}{1+\exp\left(-\dfrac{F_j-F_i}{\kappa}\right)}
\end{equation}
where $\kappa$ denotes the environmental noise coefficient, which controls the degree of randomness in the strategy updating process. When $\kappa \rightarrow 0$, the updating of the strategy becomes deterministic, and the players always mimic the strategy of the neighbor with the highest fitness. When $\kappa \rightarrow \infty$, the update of the strategy becomes completely random.

Fig.~\ref{fig:1} summarizes the evolutionary process of our model. At the initial time step, all players are randomly assigned initial strategies. In each round of the game, the evolutionary process consists of the following stages in sequential order. In the performance appraisal stage, each player is scored by its neighbors on the basis of the cooperative or defective atmosphere within the group centered on each neighbor. Subsequently, the player calculates its performance score for the current round based on the scores received from all of its direct neighbors. On this basis, the process enters the reputation update and payoff distribution stage, where players update their own reputations according to the obtained performance scores and receive their payoffs in each group based on performance allocation weight. Finally, in the strategy updating stage, players evaluate the fitness of their neighbors by combining their current-round payoffs and current reputations, and imitate the strategy of a neighbor with higher fitness with a certain probability.

\section{Simulation and results}
\label{sec:3}

In this section, Monte Carlo simulations are conducted to verify the impact of the proposed performance appraisal mechanism on the frequency of cooperation. The simulations are performed on a square lattice with periodic boundary conditions, where the network size is $N = 50 \times 50$. In the initial time step, each player is randomly assigned a strategy of cooperation (C) or defection (D) with equal probability, and the reputation values are uniformly distributed within the interval $[0, 20]$. Each evolutionary run lasts for 3000 steps. To ensure statistical stability, all results are averaged in the last 2000 steps. To eliminate the influence of stochastic fluctuations, each data point is averaged over 10 independent realizations. We set the individual contribution of a cooperator to $c = 1$. The noise coefficient is fixed at $\kappa = 0.1$.

\subsection{Effect of payoff allocation coefficient on cooperation}

\begin{figure*}
\centering 
\includegraphics[width=0.9\textwidth]{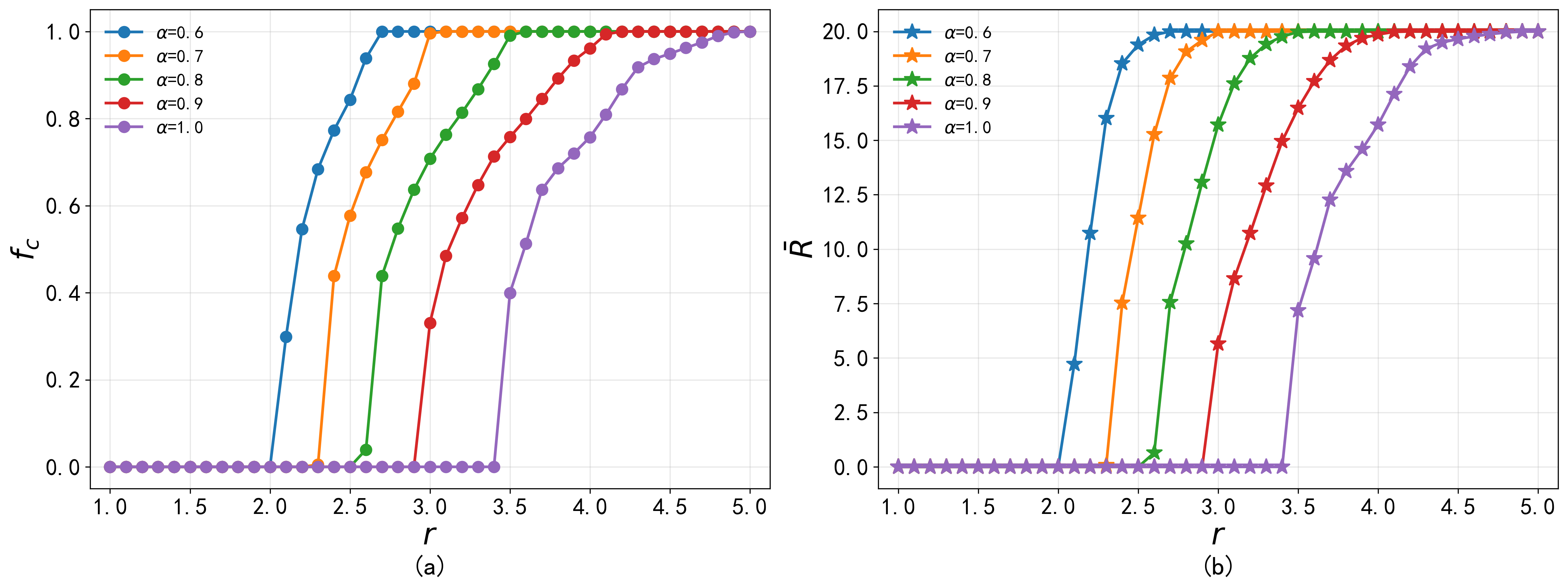}
\caption{\label{fig:2}\textbf{Cooperation frequency and average reputation as functions of the synergistic factor} $r$ 
\textbf{under different payoff distribution coefficients} $\alpha$. 
Panel (a) shows the cooperation frequency as a function of the synergistic factor $r$ for different distribution coefficients $\alpha$. Panel (b) displays the corresponding average reputation as a function of $r$. The values of $\alpha$ (0.6, 0.7, 0.8, 0.9, 1.0) are indicated in the legend. Fixed parameter $\beta = 0.5$.}
\end{figure*}

First, the payoff distribution coefficient $\alpha$ directly determines the proportions of equal distribution and performance-weighted allocation in the total payoff. We investigate the effects of different values of $\alpha$ on the evolution of cooperative behavior and system reputation.  Fig.~\ref{fig:2} presents the trends in overall cooperation frequency and average reputation as functions of the synergistic factor $r$ at different values of $\alpha$.

The results reveal a highly synchronized monotonic increasing trend between cooperation frequency and average reputation. As $r$ increases, both quantities increase rapidly and eventually converge to a state of full cooperation and the highest level of reputation. This indicates that, under the proposed allocation mechanism, the level of cooperation and the average reputation of the system exhibit a positive coevolutionary relationship.

When $\alpha = 1$, the performance-weighted proportion $(1 - \alpha)$ becomes zero and the model degenerates into the traditional public goods game with an equal payoff distribution, serving as a benchmark for comparison. In this case, the threshold for the emergence of cooperation is extremely high, with cooperative behavior occurring only when $r > 3.4$. The average reputation of the system also starts to rise slowly, only beyond this threshold. As $\alpha$ gradually decreases, the critical value of $r$ required for the emergence of cooperation is significantly reduced. When $\alpha$ decreases from 1 to 0.6, the threshold for the onset of cooperation drops from approximately 3.4 to 2.0. Meanwhile, the starting point for the increase in the average reputation also moves forward accordingly, and its growth rate becomes significantly faster, allowing the system to rapidly accumulate a high level of reputation at lower values of $r$.

The reason is that a smaller $\alpha$ corresponds to a lower proportion of equally distributed payoffs and a higher proportion of performance-weighted allocation. This mechanism generates performance weights based on neighbor evaluations, providing cooperators with consistently stronger payoff incentives. As a result, cooperative behavior can survive and spread even at lower $r$ values, i.e.,~under stronger social dilemmas. For the same value of $r$, a smaller $\alpha$ not only enhances the cooperation frequency, but also elevates the average reputation of the system simultaneously. Therefore, the performance-weighted allocation mechanism can effectively overcome the cooperation dilemma inherent in traditional equal distribution, significantly improving both the level of cooperation and the overall reputation of the system.

\subsection{Effect of reputation reinforcement coefficient on cooperation}

\begin{figure*}
\centering 
\includegraphics[width=0.9\textwidth]{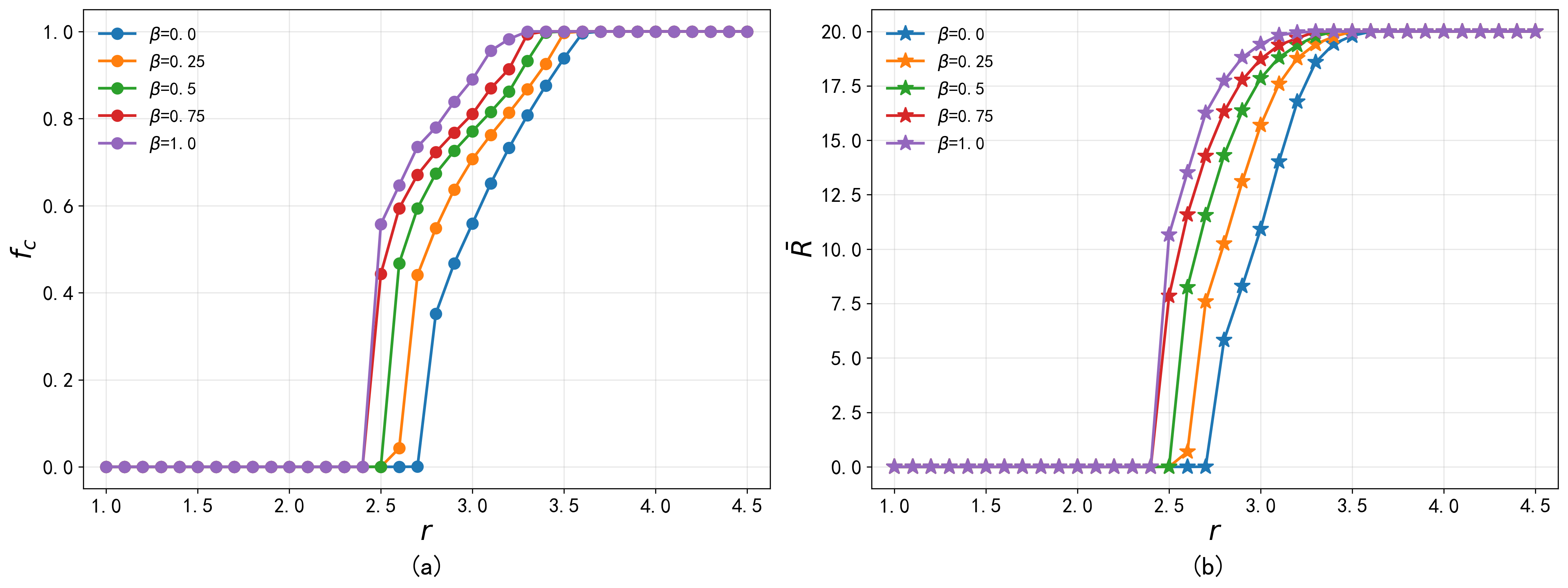}
\caption{\label{fig:3}\textbf{Cooperation frequency and average reputation as functions of the synergistic factor} $r$ \textbf{under different reputation reinforcement coefficients} $\beta$. 
Panel (a) shows the cooperation frequency as a function of $r$ for different reputation reinforcement coefficients $\beta$. Panel (b) displays the corresponding average reputation as a function of $r$. The values of $\beta$ (0, 0.25, 0.5, 0.75, 1.0) are indicated in the legend. Fixed parameter $\alpha = 0.5$.}
\end{figure*}

\begin{figure*}
\centering 
\includegraphics[width=0.9\textwidth]{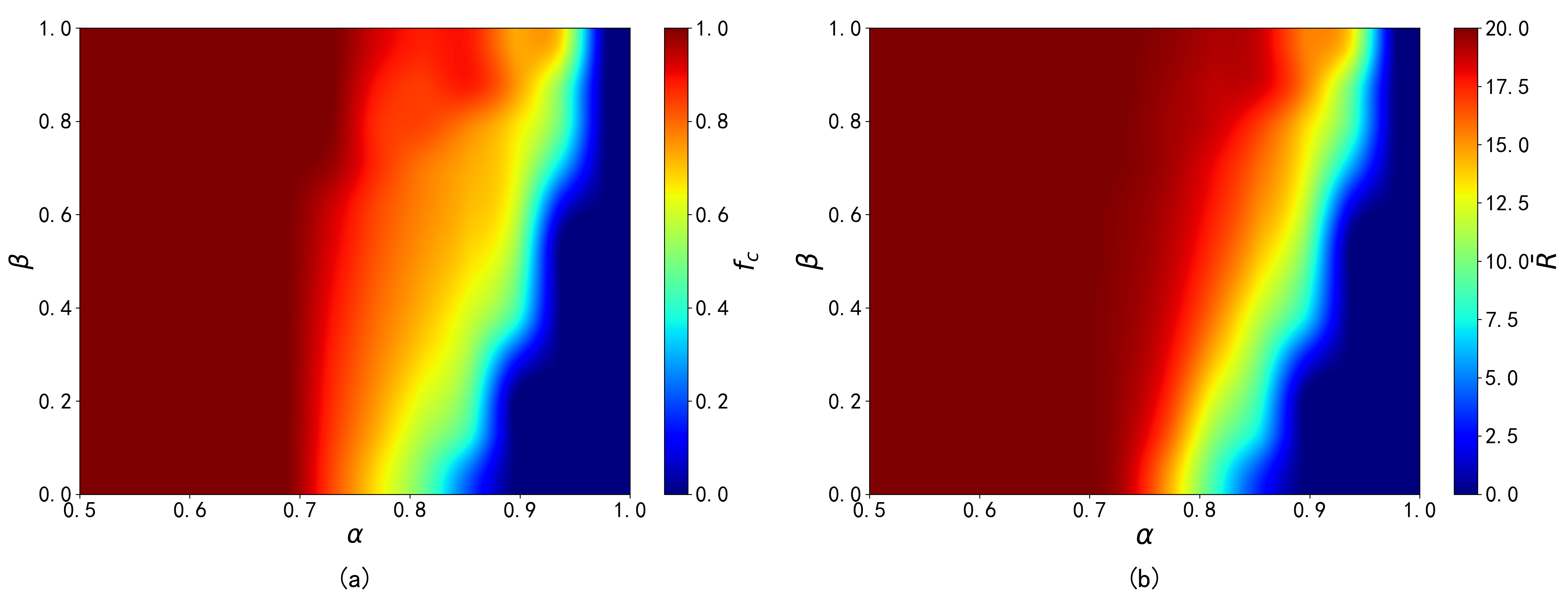}
\caption{\label{fig:4}\textbf{Joint effects of the distribution coefficient} $\alpha$ \textbf{and the reputation reinforcement coefficient} $\beta$ 
\textbf{on cooperation frequency and average reputation.}  
Panel (a) shows the distribution of cooperation frequency on the $\alpha$-$\beta$ parameter plane, and Panel (b) displays the corresponding distribution of average reputation. The color intensity indicates the level of cooperation frequency or average reputation (blue represents low values, red represents high values). The synergistic factor is fixed at $r = 3.0$. }
\end{figure*}

In this subsection, we investigate the effect of different values of $\beta$ on the evolution of cooperative behavior and system reputation. The results are presented in  Fig.~\ref{fig:3}. We find that, when other parameters are fixed, the cooperation frequency and the average reputation exhibit largely similar evolutionary trends. When $\beta = 0$, the fitness function is entirely dependent on the player's payoff, and the reputation no longer exerts a reinforcing effect on payoffs. In this case, the thresholds for the emergence of cooperation and average reputation are the highest, and cooperative behavior appears only when the synergistic factor $r > 2.7$.

As $\beta$ gradually increases, the fitness of the players begins to be influenced by reputation, and a larger $\beta$ leads to a stronger regulatory effect. Similarly, the thresholds at which cooperation frequency and average reputation start to increase decrease significantly. This implies that for the same synergistic factor $r$, a larger $\beta$ not only effectively lowers the barrier to the emergence of cooperation, but also accelerates the accumulation of the average reputation. Next, we further investigate the joint effect of the payoff distribution coefficient $\alpha$ and the reputation reinforcement coefficient $\beta$.  Fig.~\ref{fig:4}(a) presents the overall level of cooperation, and  Fig.~\ref{fig:4}(b) shows the corresponding average reputation.

As can be seen in the figures, when $\alpha$ is low, performance-weighted allocation dominates the payoff distribution. As previously discussed, this provides cooperators with immediate payoff incentives, enabling them to receive positive feedback returns in each round of the game. On this basis, the role of $\beta$ is to further translate this payoff advantage into a strategic propagation advantage. High-reputation cooperators gain greater imitation attractiveness through the fitness function, thereby accelerating the diffusion of cooperative strategies across the network. Therefore, when both parameters are in their favorable ranges, i.e.,~with low $\alpha$ and high $\beta$, the system achieves the optimal level of cooperation.

In contrast, when $\alpha$ is large, equal distribution dominates the payoff allocation. Cooperators find it difficult to obtain sufficient payoff returns through their own contributions, and thus cooperative behavior lacks a sustained incentive foundation. Under such conditions, cooperators do not possess a significant payoff advantage to begin with. Consequently, even when $\beta$ is large, the amplifying effect of reputation reinforcement on the payoffs of high-reputation cooperators remains relatively limited. As $\beta$ decreases, the reinforcing effect of reputation is further weakened, and the cooperation level continues to decline accordingly. Particularly under the parameter combination of high $\alpha$ and low $\beta$, the system falls into a dual dilemma of low cooperation and low reputation. These results indicate that when equal distribution predominates, the incentive effect of the performance appraisal mechanism is severely undermined. Reputation reinforcement can only compensate for the lack of payoff incentives to a limited extent, and cannot fundamentally reverse the trend of cooperation decline. In summary, the combination of low $\alpha$ and high $\beta$ constitutes the optimal strategy for promoting the evolution of cooperation.

\subsection{Evolution of strategy and reputation on the lattice}

\begin{figure*}
\centering 
\includegraphics[width=0.9\textwidth]{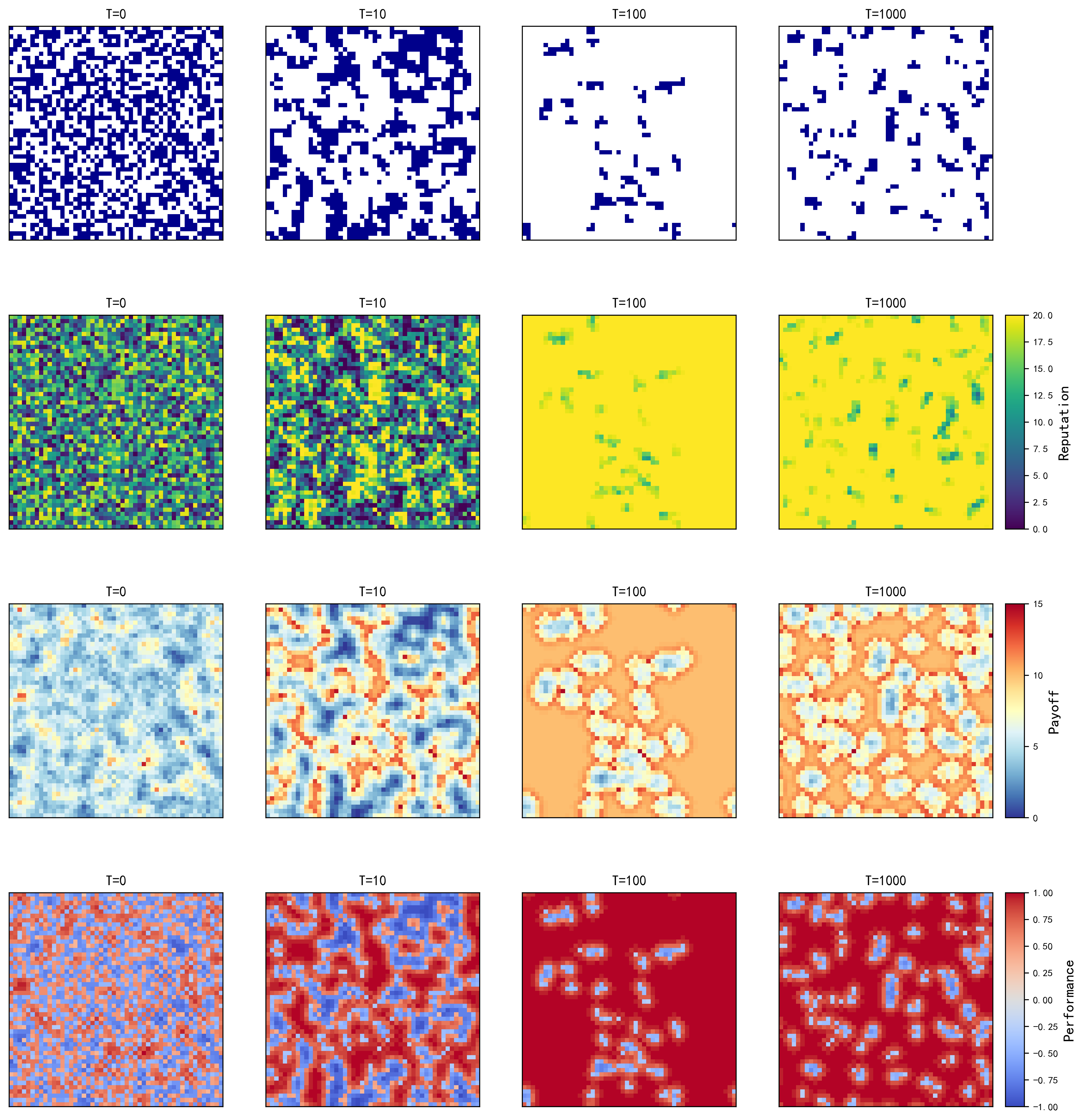}
\caption{\label{fig:5}\textbf{Spatial evolution snapshots under the emergence of cooperation.}  
The first row shows the strategy distribution at four representative time steps (blue: defectors, white: cooperators). The second row displays the corresponding reputation distribution (yellow: high reputation, purple: low reputation). The third row presents the corresponding payoff distribution (red: high payoff, blue: low payoff). The fourth row shows the corresponding performance score distribution (red: high score, blue: low score). These snapshots intuitively illustrate the strong correlations among cooperators, reputation, payoff, and scores. The parameters are set to $\alpha = 0.75$, $r = 3$, and $\beta = 0.6$. }
\end{figure*}

\begin{figure*}
\centering 
\includegraphics[width=0.9\textwidth]{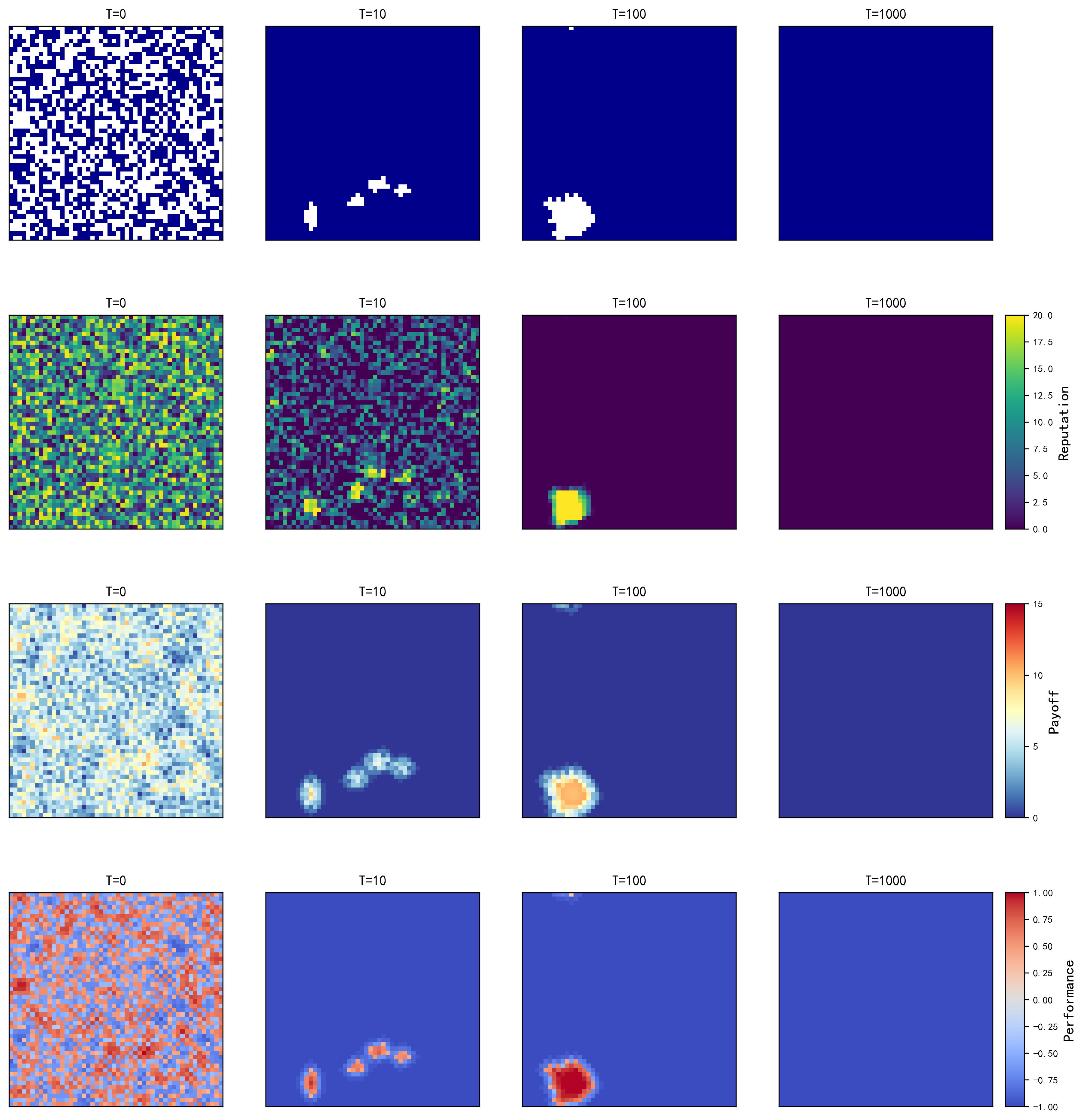}
\caption{\label{fig:6}\textbf{Spatial evolution snapshots under the collapse of cooperation.}  
The first row shows the strategy distribution at four representative time steps (blue: defectors, white: cooperators). The second row displays the corresponding reputation distribution (yellow: high reputation, purple: low reputation). The third row presents the corresponding payoff distribution (red: high payoff, blue: low payoff). The fourth row shows the corresponding performance score distribution (red: high score, blue: low score). The parameters are set to $\alpha = 0.95$, $r = 3$, and $\beta = 0.6$. }
\end{figure*}

To gain a deeper understanding of the emergence of cooperative behavior and the mechanism of formation of cooperative clusters in space, we selected a typical set of parameter combinations ($\alpha = 0.75$, $r = 3$, $\beta = 0.6$) and plotted the spatial distributions of individual strategies, reputation, payoffs, and performance scores in different time steps ($T = 0, 10, 100, 1000$), as shown in  Fig.~\ref{fig:5}.

In the initial time step ($T = 0$), the strategies, reputations, payoffs, and performance scores of the individuals are all completely randomly distributed. When evolution progresses to $T = 10$, cooperators gradually aggregate into clusters in space. From the performance score distribution, it can be observed that cooperators inside the clusters have higher scores, while boundary cooperators adjacent to defectors have slightly lower scores than their interior counterparts. This score gradient provides interior cooperators with an advantage in the payoff distribution, thereby creating conditions for the further expansion of cooperators.

By $T = 100$, cooperator clusters have further expanded and gradually eroded the living space of defectors. At this stage, some small defector clusters are surrounded by cooperators. Although these besieged defectors lower the performance scores of adjacent cooperators, their own scores are higher than those of defectors located deep inside defector clusters, thereby narrowing the payoff gap between them and boundary cooperators to some extent. When evolution progresses to $T = 1000$, a small number of defectors begin to appear within some large clusters of cooperators. These defectors survive by hiding within cooperator clusters and taking advantage of free-riding behavior.

To further reveal the underlying reasons why the performance appraisal mechanism promotes the emergence of cooperation, we selected a larger value of $\alpha$ ($\alpha = 0.95$) and conducted a comparative analysis with the aforementioned case of $\alpha = 0.75$, thus examining the impact of the payoff distribution on the evolution of cooperation, as shown in  Fig.~\ref{fig:6}.

Under the parameter setting of $\alpha = 0.95$, the strategy distribution at the initial time step ($T = 0$) remains a randomly mixed state. By $T = 10$, the cooperators have been compressed into a few small scattered and isolated clusters. By $T = 100$, only one cooperator cluster has achieved limited expansion. As evolution progresses to $T = 1000$, cooperators completely disappear from the system.

Further analysis of the remaining clusters of cooperators in $T = 10$ and $T = 100$ reveals that, although these cooperators possess higher reputation and performance scores than defectors, no significant payoff gap has formed between them and the surrounding defectors. More critically, due to the excessively small size of the cooperator clusters, they lack internal high-payoff cooperators to support the boundary cooperators, rendering these clusters highly unstable and vulnerable to invasion by defectors.

The above results verify the key conditions under which cooperator clusters can successfully resist invasion by defectors: only when $\alpha$ takes a sufficiently small value and a payoff gradient is formed between the internal high-payoff layer and the boundary layer within the cooperator cluster can the performance appraisal mechanism effectively take effect and drive the existence and sustained expansion of cooperative behavior.

%%%%%%%%%%%%%%%%%%%%%%%%%%%%%%%%%%%%%%%%
\subsection{Robustness analysis on different network topologies}

\begin{figure*}
\centering 
\includegraphics[width=0.9\textwidth]{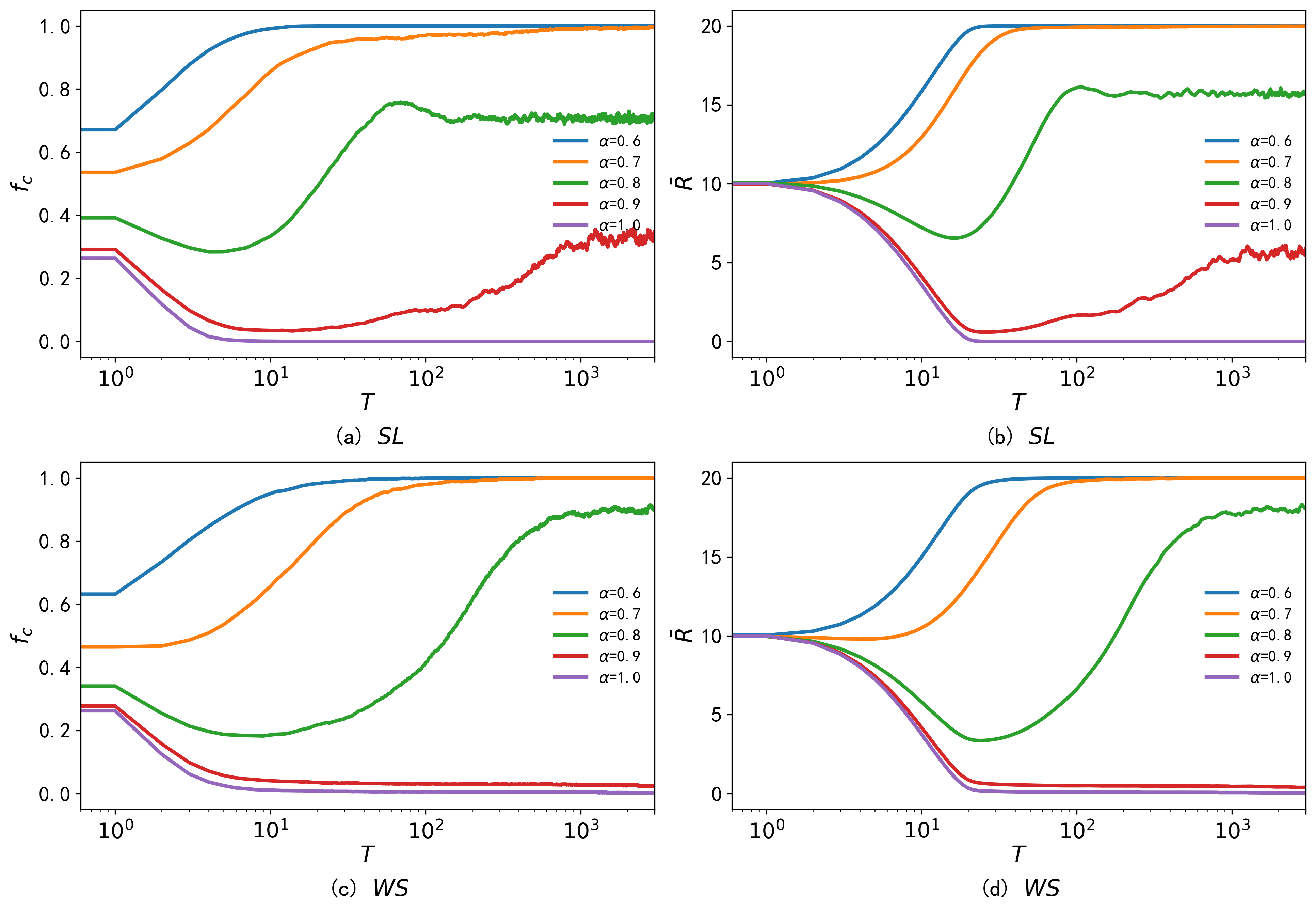}
\caption{\label{fig:7}\textbf{Temporal evolution trajectories of cooperation frequency and average reputation under different distribution coefficients} $\alpha$ \textbf{on square lattice and small-world networks.}  
Different colored curves represent different values of the distribution coefficient $\alpha$, as indicated in the legend. Panels (a) and (b) show the evolution trajectories of cooperation frequency and average reputation as functions of Monte Carlo steps on the square lattice, respectively. Panels (c) and (d) show the corresponding evolution trajectories on the small-world network (with rewiring probability $p = 0.1$), respectively. The parameters are set to $r = 3$ and $\beta = 0.25$. }
\end{figure*}

To further verify the robustness of the observations obtained under the performance appraisal mechanism, we compare the temporal evolution processes on the square lattice and the small-world network. The square lattice only allows for local interactions, while the small-world network introduces a probability of rewiring of $p = 0.1$.  Fig.~\ref{fig:7} shows the temporal evolution trajectories of the cooperation frequency and average reputation for different distribution coefficients $\alpha$, under the parameters $r = 3$ and $\beta = 0.25$.

When $\alpha = 1$, the entire payoff of the players comes from the same distribution and the performance-weighted allocation proportion is zero. This setting favors defectors, since free-riding behavior is rapidly imitated. Consequently, the defection eventually spreads throughout the network, and the average reputation drops to its minimum.

When $\alpha = 0.9$, the majority of payoffs come from an equal distribution, with only a small portion derived from performance-weighted allocation. In the square lattice network, both the cooperation frequency and the average reputation first decline and then increase, eventually converging to stable values. However, in the small-world network, the evolution of cooperation fails. When $\alpha = 0.8$, regardless of whether the network is a square lattice or small-world, the system ultimately reaches a stable state where cooperators and defectors coexist in a fixed ratio. When $\alpha = 0.7$ or $0.6$, the performance-weighted allocation proportion increases, allowing cooperators to achieve rapid payoff growth. Under both network topologies, the cooperation frequency and the average reputation increase rapidly, and ultimately cooperators occupy the entire network.

These results demonstrate that the proposed performance-weighted allocation mechanism effectively promotes cooperation across different network topologies. By adjusting the value of $\alpha$, cooperative behavior is gradually encouraged, thereby indicating its robustness across different network structures.

%%%%%%%%%%%%%%%%%%%%%%%%%%%%%%%%%
\section{Conclusions and outlook}
\label{sec:4}

In this paper, we extend the traditional SPGG by proposing a performance appraisal payoff distribution mechanism. This mechanism is designed to capture a fundamental feature of wage distribution in the real world: workers' compensation typically consists of a base salary and a performance component, where the latter is linked to the individual's daily performance score, effectively incentivizing employees to consistently and reliably fulfill their duties. In our model, the performance score of a player is evaluated by its neighbors based on the cooperative or defective atmosphere of their groups, thereby enabling dynamic scoring of players. Furthermore, we assume that players' strategy learning is associated with their reputation, and players with a higher reputation achieve greater fitness through the fitness function, thereby enhancing the propagation advantage of their strategies.

Our findings reveal that a smaller distribution coefficient $\alpha$, which corresponds to a higher proportion of performance-weighted allocation, is conducive to incentivizing cooperators and thus promoting the evolution of cooperation. This observation is highly consistent with real-world organizational management practices: in firms or teams where performance pay constitutes a larger share of total compensation, employees tend to exhibit more proactive contribution behaviors, and free-riding is relatively less common. In other words, when individual income is closely tied to actual performance, cooperative behavior is more likely to spontaneously emerge in competitive environments. 

Furthermore, by incorporating reputation into the fitness function, we grant high-reputation individuals a greater fitness advantage in strategy imitation, which mirrors a widespread mechanism in human societies: individuals with a good reputation are more likely to gain trust, resources, and social influence, thus further reinforcing their cooperative behavior. The joint effect of $\alpha$ and $\beta$ reveals a synergistic interaction in promoting cooperation. The simulation results show that when $\alpha$ is low and $\beta$ is high, the system achieves the highest level of cooperation. In this case, cooperators have a significant payoff advantage, and a larger $\beta$ further amplifies this advantage, enabling high-reputation cooperators to spread their strategies more rapidly. In contrast, when $\alpha$ is large, even a high $\beta$ fails to significantly improve the cooperation level. This is because the payoff distribution is dominated by equal sharing, leaving cooperators without a substantial payoff advantage, and thus the amplifying effect of $\beta$ loses its functional basis. This indicates that $\alpha$ serves as a fundamental factor determining the survival of cooperation: only when performance-based allocation dominates can cooperators obtain immediate payoff incentives that sustain their existence. The role of $\beta$, in turn, is to amplify this incentive effect by accelerating the diffusion of cooperative strategies through the reputation mechanism.

Additionally, we have validated the robustness of our model across different network topologies and found that the proposed mechanism effectively promotes cooperation in all tested structures. Therefore, the mechanism proposed in this paper exhibits a certain degree of structural universality and provides theoretical insight for the design of incentive systems in the real world.

In our work, an individual's score is entirely determined by the evaluations of its neighbors. However, in the real world,
the performance appraisal criteria are often more diverse. Future research could take into account the past work performance
of employees~\cite{lu2025past,pi2022evolutionary} or introduce additional evaluation indicators to establish more
comprehensive assessment standards. In addition, random perturbations could be incorporated into the scoring process to
simulate evaluation noise and uncertainty in the real world~\cite{fujimoto2023evolutionary,hilbe2018indirect}, further
enhancing the ecological validity and applicability of the model.

%%%%%%%%%%%%%%%%%%%%%%
\section*{Acknowledgments}

T. Li, Q. Li, K. Zhu, and M. Feng are supported by the Fundamental Research Funds for the Central Universities (Grant No. SWU-KT26010) and the Natural Science Foundation of Chongqing, China (Grant No. CSTB2025YITP-QCRCX0007). M. Chica was supported by the Andalusian Government under grant Emergia (EMERGIA21\_00139) and by grant CNS2024-154265 funded by MCIN/AEI/ 10.13039/501100011033 and, as appropriate, by "ESF Investing in your future" or by "European Union NextGenerationEU/PRTR".

\bibliographystyle{unsrt}
% \bibliographystyle{cas-model2-names}

\bibliography{cas-refs}



\end{document}